\documentstyle[12pt,moriond,graphics,vcaption2]{article}
\newcommand{\lya}{Ly$\alpha$}
\newcommand{\bpar}{$b$-parameter}
\newcommand{\cm}{\rm cm}
\newcommand{\K}{\rm K}

\newcommand{\ls}{\mathrel{\mathchoice{\vcenter{\offinterlineskip\halign{\hfil
$\displaystyle##$\hfil\cr<\cr\sim\cr}}}
{\vcenter{\offinterlineskip\halign{\hfil$\textstyle##$\hfil\cr<\cr\sim\cr}}}
{\vcenter{\offinterlineskip\halign{\hfil$\scriptstyle##$\hfil\cr<\cr\sim\cr}}}
{\vcenter{\offinterlineskip\halign{\hfil$\scriptscriptstyle##$\hfil\cr<\cr\sim\cr}}}}}
\newcommand{\gs}{\mathrel{\mathchoice{\vcenter{\offinterlineskip\halign{\hfil
$\displaystyle##$\hfil\cr>\cr\sim\cr}}}
{\vcenter{\offinterlineskip\halign{\hfil$\textstyle##$\hfil\cr>\cr\sim\cr}}}
{\vcenter{\offinterlineskip\halign{\hfil$\scriptstyle##$\hfil\cr>\cr\sim\cr}}}
{\vcenter{\offinterlineskip\halign{\hfil$\scriptscriptstyle##$\hfil\cr>\cr\sim\cr}}}}}

\begin{document}
\heading{MEASURING THE TEMPERATURE OF THE INTERGALACTIC MEDIUM}

\author{Joop Schaye $^{1}$, Tom Theuns $^{2,1}$, Anthony Leonard
$^{3}$, George Efstathiou $^{1}$}
{$^{1}$ Institute of Astronomy, Cambridge, UK.}  
{$^{2}$ Max-Planck-Institut f\"ur Astrophysik, Garching, Germany.} 
{$^{3}$ Department of Physics, Astrophysics, Oxford, UK.} 

\begin{moriondabstract}
Numerical simulations indicate that the smooth, photoionized
intergalactic medium (IGM) responsible for the low column density
\lya\ forest follows a well defined temperature-density relation. We
show that such an equation of state results in a cutoff in the
distribution of line widths ($b$-parameters) as a function of column
density ($N$) for the low column density absorption lines. This
explains the existence of the lower envelope which is clearly seen in
scatter plots of the $b(N)$-distribution in observed QSO spectra. The
intercept and slope of this cutoff can be used to measure the equation
of state of the IGM. Measuring the evolution of the equation of state
with redshift will allow us to put tight constraints on the
reionization history of the universe.
\end{moriondabstract}

\section{Introduction}
Neutral hydrogen in the intergalactic medium (IGM) gives rise to a
forest of \lya\ absorption lines blueward of the \lya\ emission line
in quasar spectra.  Hydrodynamic simulations of structure formation in
a universe dominated by cold dark matter and including an ionizing
background, show that the low column density ($N \ls
10^{14.5}\,\cm^{-2}$) \lya\ forest at redshift $z \gs 2$ is produced
by a smoothly varying IGM. For the low-density gas responsible for the
\lya\ forest, shock heating is not important and the gas follows a
tight temperature-density relation. The competition between
photoionization heating and adiabatic cooling results in a power-law
`equation of state' $T=T_0(\rho/\bar{\rho})^{\gamma-1}$, which depends
on cosmology and reionization history~\cite{hui97:tempdens}. 

The smoothly varying IGM gives rise to a fluctuating optical depth in
redshift space. Many of the optical depth maxima can be fitted quite
accurately with Voigt profiles. The distribution of line widths
depends on the initial power spectrum, the peculiar velocity gradients
around the density peaks and on the temperature of the IGM. However,
there is a lower limit to how narrow the absorption lines can
be. Indeed, the optical depth will be smoothed on a scale determined
by three processes: thermal broadening, baryon (Jeans) smoothing and
possibly instrumental, or in the case of simulations, numerical
resolution. The first two depend on the thermal state of the
gas. While for high-resolution observations (echelle spectroscopy) the
effective smoothing scale is not determined by the instrumental
resolution, numerical resolution has in fact been the limiting factor
in many simulations (see~\cite{theuns98:apmsph} for a discussion).

Scatter plots of the $b(N)$-distribution have been published for many
observed QSO spectra
\cite{hu95:lyaobs,lu96:lya_z=4,kirkman97:lya_obs,kim97:lya_evolution}.
These plots show a clear cutoff at low $b$-parameters, which increases
slightly with column density. However, this cutoff is not absolute,
there are some narrow lines, especially at low column densities. Lu et
al.~\cite{lu96:lya_z=4} and Kirkman \& Tytler~\cite{kirkman97:lya_obs}
use Monte Carlo simulations to show that many of these lines are
caused by line blending and noise in the data. Some contamination from
unidentified metal lines is also expected. A lower envelope which
increases with column density has also been seen in numerical
simulations \cite{zhang97:lya_forest}.

In this contribution we shall demonstrate that the cutoff in the
$b(N)$-distribution is determined by the equation of state of the
low-density gas and can therefore be used to measure the equation of
state of the IGM. This work will be more fully described and discussed
in a forthcoming publication~\cite{schaye99:eos}.

\section{Results}

\begin{figure*}
\resizebox{\textwidth}{!}{\includegraphics{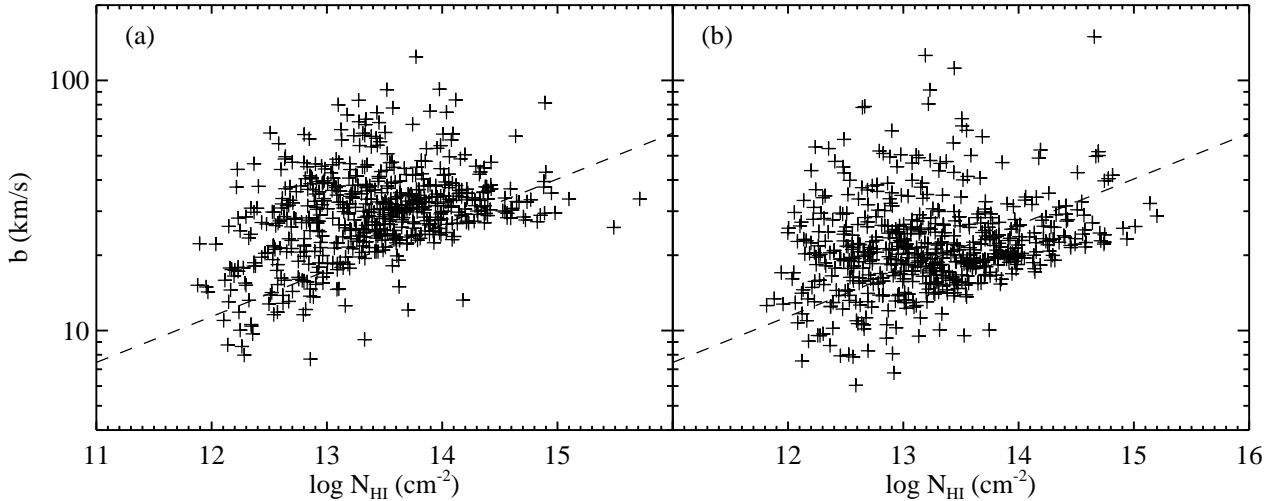}}
\caption{The $b(N)$-distribution for 800 random lines from models with
$\log T_0 = 4.20$ (a) and $\log T_0 = 3.83$ (b) respectively, at
$z=3$. The position of each line is 
indicated by a cross. Errors are not displayed. The dashed line is the
cutoff for the lines plotted in panel (a) over the range $10^{12.5}\,\cm^{-2}
\le N \le 10^{14.5}\,\cm^{-2}$.}  
\label{fig:bN}
\end{figure*}

In Fig.~\ref{fig:bN}a we plot the $b(N)$-distribution for 800 random
absorption lines taken from multicomponent Voigt profile fits of
spectra at redshift $z=3$, generated from one of our simulated
models. A cutoff at low $b$-values, which increases with column
density, can clearly be seen. As in the observations, there are some
very narrow lines, which occur in blends. These unphysically narrow
lines make determining the cutoff in an objective manner
nontrivial. We developed an iterative procedure for fitting a
power-law, $b = b_0 (N/N_0)^{\Gamma -1}$, to the $b(N)$-cutoff over a
certain column density range ($10^{12.5}\,\cm^{-2} \le N \le
10^{14.5}\,\cm^{-2}$ at $z=3$) which is insensitive to these narrow
lines.

The $b(N)$-distribution for a
colder ($\log T_0 = 3.83$ vs.\ $\log T_0 = 4.20$) model is plotted in
Fig.~\ref{fig:bN}b. Clearly, the distribution cuts off at lower
$b$-values. Let us assume that the absence of lines with low
$b$-values is due to the fact that there is a minimum line width set
by the thermal state of the gas through the thermal broadening and/or
baryon smoothing scales. Since the temperature of the low-density gas
responsible for the \lya\ forest increases with density, we expect the
minimum $b$-value to increase with column density, provided that the
column density correlates with the density of the absorber.

\begin{figure}
\resizebox{0.5\textwidth}{!}{\includegraphics{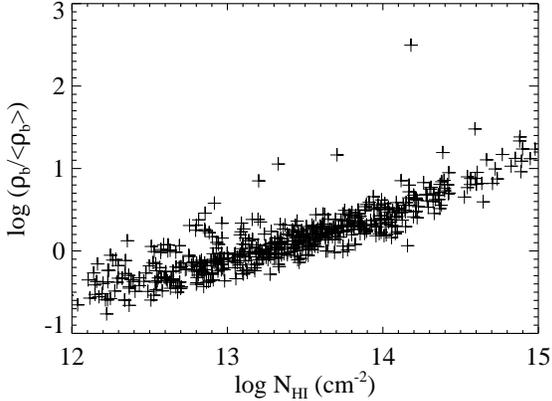}}
\caption{The gas density at the line centres as a
function of neutral hydrogen column density. Data points correspond to
the lines plotted in Fig.~\ref{fig:bN}a.}
\label{fig:dens-N}
\end{figure}

To see whether this picture is correct, we need to investigate the
relation between the Voigt profile parameters $N$ and $b$, and the
density and temperature of the absorbing gas respectively.  From
Fig.~\ref{fig:dens-N}, which shows the gas density as a function of
column density for the absorption lines plotted in Fig.~\ref{fig:bN}a,
it can be seen that these two quantities are tightly correlated.

\begin{figure*}
\resizebox{\textwidth}{!}{\includegraphics{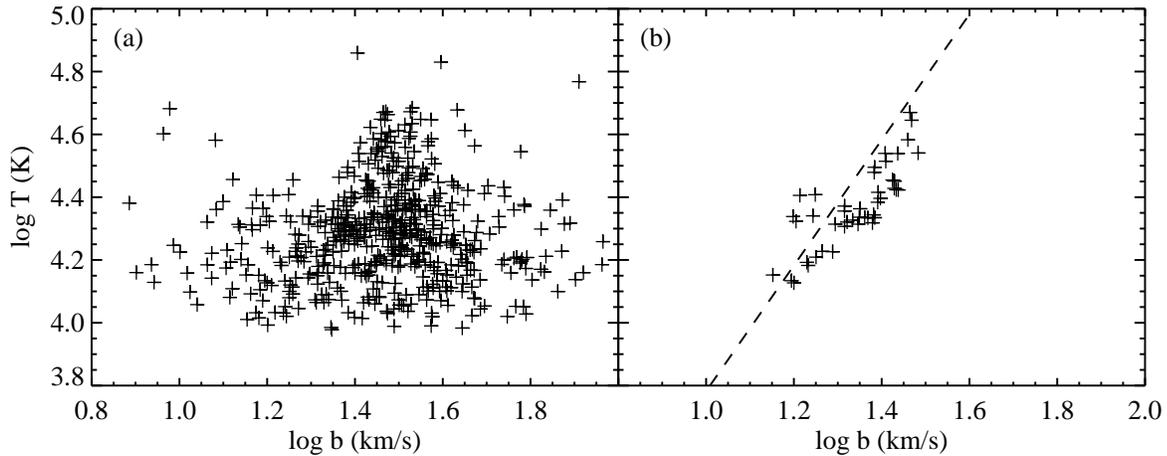}}
\caption{The temperature at the line centres as
a function of \bpar. The left panel contains all lines plotted in
Fig.~\ref{fig:bN}a. In the right panel only those lines are 
plotted which have \bpar s within one mean absolute deviation of the
power-law fit to the $b(N)$-cutoff plotted as the dashed line in
Fig.~\ref{fig:bN}a. The dashed line corresponds to the thermal width,
$b=(2 k_B T/m_p)^{1/2}$.}
\label{fig:temp-b}
\end{figure*}

The temperature is plotted against the $b$-parameter in
Fig.~\ref{fig:temp-b}a. The result is a scatter plot with no apparent
correlation. This is not surprising since many absorbers will be
intrinsically broader than the local thermal broadening scale. In
order to test whether the cutoff in the $b(N)$-distribution is a
consequence of the existence of a minimum line width set by the
thermal state of the gas, we need to look for a correlation between
the temperature and \bpar s of the lines near the
cutoff. Fig.~\ref{fig:temp-b}b shows that these lines do indeed
display a tight correlation. The dashed line corresponds to the
thermal width, $b=(2 k_B T/m_p)^{1/2}$, where $m_p$ is the mass of a
proton and $k_B$ is the Boltzmann constant. Lines corresponding to
density peaks whose width in velocity space is much smaller than the
thermal broadening width, have Voigt profiles with this \bpar .

Figs.\ \ref{fig:dens-N} and \ref{fig:temp-b}b suggest that the cutoff
in the $b(N)$-distribution should be strongly correlated with the
equation of state of the absorbing gas.  The objective is to establish
the relations between the cutoff parameters and the equation of state
using simulations. These relations turn out to be unaffected by
systematics like changes in cosmology (for a fixed equation of
state)~\cite{schaye99:eos} and can thus be used to measure the 
equation of state of the IGM using the cutoff in the observed
$b(N)$-distribution.

\begin{center}
\begin{figure}
\resizebox{\textwidth}{!}{\includegraphics{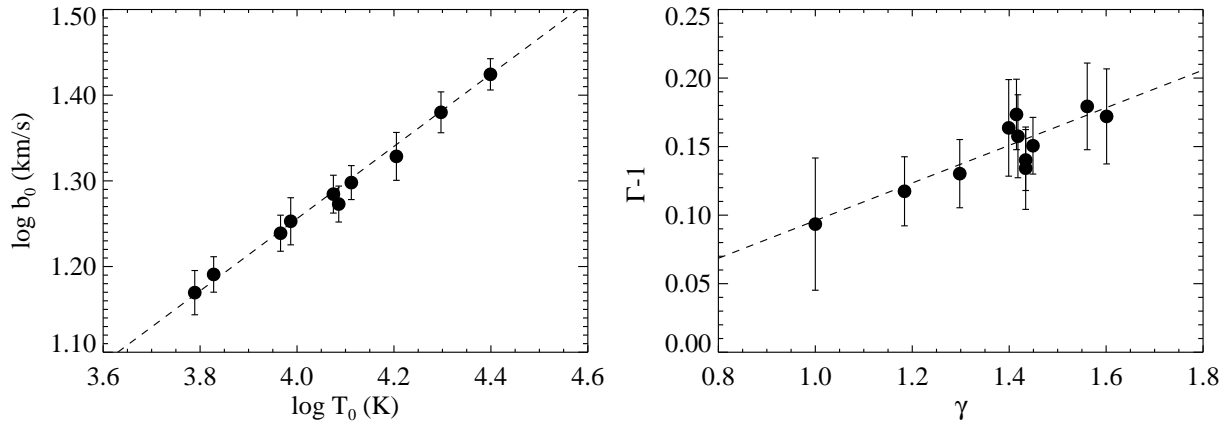}}
\caption{The intercept of the $b(N)$-cutoff as a function of the temperature
at mean density (left panel) and the slope as a function of the index of the
power-law equation of state (right panel), for redshift $z=3$. The
error bars enclose 68\% 
confidence intervals around the medians, as determined from 500
sets of 300 random lines. The dashed
lines are the maximum likelihood fits.}
\label{fig:relations}
\end{figure}
\end{center}

The amplitudes of the power-law fits to the cutoff and the equation of
state are plotted against each other in the left panel of
Fig.~\ref{fig:relations}.  The error bars, which indicate the
dispersion in the cutoff of sets of 300 lines (typical for $z=3$), are
small compared to the differences between the models. This means that
measuring the cutoff in a single QSO spectrum can provide significant
constraints on theoretical models (at $z=3$, physically reasonable
ranges for the parameters of the equation of state are $10^{3.0}\,\K <
T_0 < 10^{4.5}\,K$ and $1.2 < \gamma < 1.7$
\cite{hui97:zeldovich_coldensdistr,hui97:tempdens}).  The slope of the
cutoff, $\Gamma-1$, is plotted against $\gamma$ in the right panel of
Fig.~\ref{fig:relations}. The dispersion in the slope of the cutoff
for a fixed equation of state is comparable to the difference between
the models. The weak dependence of $\Gamma$ on $\gamma$ and the large
spread in the measured $\Gamma$ will make it difficult to put tight
constraints on the slope of the equation of state.


%
\vfill
\end{document}